# Feasibility of Lithium Storage on Graphene and Its Derivatives

*Yuanyue Liu, † Vasilii I. Artyukhov,† Mingjie Liu,† Avetik R. Harutyunyan‡ and Boris I. Yakobson†\**

† Department of Mechanical Engineering and Materials Science, Department of Chemistry, and the Smalley Institute for Nanoscale Science and Technology, Rice University, Houston, Texas, 77005, USA

‡ Honda Research Institute USA, Inc., Columbus, Ohio, 43212, USA

ABSTRACT: Nanomaterials are anticipated to be promising storage media, owing to their high surface-to-mass ratio. The high hydrogen capacity achieved by using graphene has reinforced this opinion and motivated investigations of the possibility to use it to store another important energy carrier – lithium (Li). While the first-principles computations show that the Li capacity of pristine graphene, limited by Li clustering and phase separation, is lower than that offered by Li intercalation in graphite, we explore the feasibility of modifying graphene for better Li storage. It is found that certain structural defects in graphene can bind Li stably, yet more efficacious approach is through substitution doping with boron (B). In particular, the layered $C_3B$ compound stands out as a promising Li storage medium. The monolayer $C_3B$ has a capacity of 714 mAh/g (as $Li_{1.25}C_3B$), and the capacity of stacked $C_3B$ is 857 mAh/g (as $Li_{1.5}C_3B$), which is about twice as large as graphite's 372 mAh/g (as $LiC_6$). Our results help clarify the mechanism of Li storage in low-dimensional materials, and shed light on the rational design of nano-architectures for energy storage.

The search for high energy density electrodes is one of the central topics in lithium (Li) ion battery studies.[1-6] The energy density is proportional to the product of full-cell voltage times Li capacity.[3] Nano-materials have been expected to have high storage capacities due to their high surface-to-mass ratio, as compared to three-dimensional (3D) bulk materials. For example, two-dimensional (2D) carbon -- graphene, with its record surface-to-mass ratio of 2630 m$^2$/g, has proven to be a promising matrix for hydrogen storage.[7-11] However, the experimental studies of Li storage on graphene remain controversial, and it is still not clear whether graphene could have a higher capacity than graphite, which is used commercially as an anode with a capacity of 372 mAh/g (340 mAh/g, including Li own weight). Some experiments do show high Li capacity for graphene nano-sheets, within a few charge/discharge cycles.[12-16] Yet detailed examination of graphene quality attributes the Li storage to binding with defects, which are created during the fabrication of nano-sheets.[17, 18] Furthermore, *in situ* Raman spectroscopy indicates that the amount of Li absorbed on monolayer graphene is greatly reduced compared to graphite, while the intercalation of Li into few layer graphene seems to resemble that of graphite.[19] In order to further clarify this issue, we perform first-principles computations to assess the Li storage in the carbon (C) based nano-materials. We start from the general description of obtaining battery characteristics from calculations, and then apply it to Li-graphene system, which shows a distinguishing Li storage behavior compared with graphite. The feasibility of modifying graphene for the Li storage is further explored, which leads to the finding that the layered $C_3B$ compound could be a promising storage medium.

The materials used as electrodes for Li storage should have binding strength with Li within certain range. On the one hand, binding to anode material matrix (M) should be weaker than on the cathode side, to ensure the chemical potential driving force for subsequent Li migration from anode to cathode during discharge; this binding energy difference divided by electron charge $e$ gives the average discharge voltage.[3, 20] On the other hand, this binding energy $\varepsilon_{Li-M}$ should be greater than cohesive energy $\varepsilon_{Li}$ of bulk Li, in order to prevent phase separation and formation of hazardous Li dendrites.[1] The theoretical capacity of the matrix (M) is determined by the highest Li:M ratio (commonly expressed in the units of mAh/g) that can be achieved in the stable compounds without phase separation, i.e. Li precipitation on the anode. Generally, it can be found by considering the energy $\varepsilon(x`)$ per *average atom* in the composition $Li_{x`}M_{1-x`}$, $\varepsilon(x`)$ curve.[21] In context of electrode, since the matrix M essentially retains its fixed amount, here we find it more convenient to determine the capacity from the lithiation energy E per *matrix unit* versus composition variable $x$ in $Li_xM$, a lithiation curve $E(x)$ defined as:

$$E(x) \equiv E(Li_xM) + x \cdot \varepsilon_{Li} - E(M), \qquad (1)$$

where $E(Li_xM)$ is the energy of $Li_xM$ and $E(M)$ is the energy of matrix M, both w.r.t atomic states of constituent elements. The number of M atoms in the matrix unit, can be normalized to 6 atoms, so that graphite's known charged phase would have $x = 1$, which allows for a convenient comparison of capacities for different matrix materials. A number of physical quantities can be extracted from the lithiation curve. First, according to Equation 1, the lithium–matrix binding energy $\varepsilon_{Li-M}$, relative to the cohesive energy $\varepsilon_{Li}$ of bulk Li, can be determined from the curves as $\varepsilon_{Li-M} - \varepsilon_{Li} = -E(x)/x$ (and is linearly related to the average discharge voltage, as mentioned above). Second, following the basic thermodynamics definitions, the value of Li chemical potential (again, relative to the bulk Li, and neglecting temperature and entropy effects) is simply a derivative of the lithiation curve, $\partial E(x)/\partial x$. Thus, a negative slope of the lithiation curve suggests that more Li can be stored, while a positive slope means that Li would rather precipitate from that composition, leading to the phase separation and the formation of dendrites. Therefore, the achievable capacity limit is determined by the position $x$ of the minimum of the $E(x)$ curve (possibly with some excess permitted by the nucleation barrier to the Li precipitation). We obtain the $E(x)$ plots by first-principles computations, assisted by the cluster expansion method.[22, 23] The detailed description of calculations can be found in Supporting Information (SI). Representative points from the full lithiation curves (shown in SI) are plotted in Figure 1 and 3. These points correspond to the ground-state configurations at each respective composition. The solid circles mark the Li-saturated (fully charged) phases, while the continued dashed curves show the concentration ranges prone to metallic Li precipitation.

In Figure 1, the *graphite* lithiation curve is negative with a minimum at $x = 1$, corresponding to a stable compound, $LiC_6$, with a capacity of 372 mAh/g, in agreement with the literature.[1] The atomic structure of $LiC_6$ is shown in Figure 1 as well, where the numbers (in eV) are the energy cost for adding (or removing) a single Li atom to (or from) bulk $LiC_6$ of large size. All the numbers are positive, indicating that the compound is indeed stable.

For *graphene*, in contrast, the lithiation energy in Figure 1 is always positive, monotonically increasing with Li loading, indicating that the capacity is, in fact, zero! The contrasting lithiation behaviors result from the different $\varepsilon_{Li-M}$. Although in both cases Li loses its 2$s$ electron to C, producing ionic Li–C bonding, the bonding energies are different: $\varepsilon_{Li-graphite} > \varepsilon_{Li} > \varepsilon_{Li-graphene}$. For example, at $x = 1$, $\varepsilon_{Li-graphite} - \varepsilon_{Li} = 0.07$ eV, while $\varepsilon_{Li-graphene} - \varepsilon_{Li} = -0.61$ eV. Therefore, when loaded with Li, the energy of Li–graphite system drops due to the increase in the favorable Li–graphite bonding, until reaching the $LiC_6$ composition, where further Li loading results in a strong repulsion between Li ions at neighboring hexagons.[24] In contrast, the energy of Li−graphene system rises during lithiation due to the increasing amount of relatively unfavorable Li−graphene interactions, accompanied by the Li−Li ions repulsion. Moreover, the positive lithiation energy of graphene means that the Li adatoms on it should aggregate into clusters and eventually macroscopic dendrites, instead of forming any stable Li−graphene mixture phase.

Why does the $\varepsilon_{Li\text{-}M}$ differ so much between graphite and graphene? In graphite, the Li ions are intercalated between two C layers, while on graphene, the Li ions are only adsorbed on surface. The intercalation configuration raises the $\varepsilon_{Li\text{-}M}$ due to the increased Li coordination (greater "contact area" with the matrix). The role of intercalation is further evident in the lithiation of bilayer, as shown in Figure 1. Our calculations show that it is energetically favorable for the Li ions to enter between the C layers, rather than to be adsorbed on the exterior surface. Due to the available intercalation sites, bilayer graphene can store Li in the form of $LiC_{16}$. Another nearly-degenerate in energy form $LiC_{12}$ is also found, with the Li ordering between two layers similar to that in graphite, Figure 1 (energy difference being only ~2 meV, which is within calculation accuracy; proper treatment of van der Waals interactions might help distinguish their energies.[25]). The $\varepsilon_{Li\text{-}bilayer}$ is close to the $\varepsilon_{Li\text{-}graphite}$ at the corresponding Li-saturated configurations, with the former binding slightly stronger by 0.06 eV/Li, indicating again that the enhanced binding is mainly due to the intercalation configuration. In summary, although graphene (monolayer or multilayer) provides more accessible surface area, the exposed surfaces turn relatively inactive, with Li binding weak, which is unable to prevent Li phase separation, and consequently leads to a reduced capacity.

However, the accessibility of the open graphene forms and almost certainly faster surface diffusion are very attractive for better kinetic performance of the electrodes. To remedy the insufficient binding, graphene surfaces might be "activated" by several means briefly assessed below.

*Elastic deformation.* One can reasonably hypothesize that curvature of graphene lattice should change purely $sp^2$-hybridization to partially $sp^3$ (often quantified by the pyramidalization angle),[26, 27] making C lattice more chemically active. To evaluate this possibility, we have computed the binding energies, to show in Figure 2 how the purely elastic curvature of carbon nanotube (CNT) wall enhances binding with a single Li atom. As the diameter increases, the $\varepsilon_{Li-CNT}$ decreases and asymptotically approaches the $\varepsilon_{Li-graphene}$. Interestingly, the single Li atom prefers adsorption on the outer rather than the inner surface of CNT wall (though the difference is small, < 0.03 eV), while at high Li concentrations, the inner surfaces become more favorable than the outer. However, for small-diameter CNT such as (5, 5), the energy preferences are reversed. While any systematic investigation of elastic curvature effects on binding strength is beyond the scope of this study, several computed samples are already informative. In all cases, the $\varepsilon_{Li-CNT}$ is still less than the cohesive energy $\varepsilon_{Li}$ of bulk Li, which indicates that the single-wall CNT cannot form stable compound with Li and thus has low capacity.

*Native structural disorder,* such as pentagons, heptagons, dislocations, Stone-Wales defects, mono- or di-vacancies, ad-dimers, and edges. Figure 2 shows the configurations of Li complexes with such defects, and the relative binding energies, $\varepsilon_{Li} - \varepsilon_{Li-defect}$. While pristine graphene cannot effectively adsorb a single Li atom from its bulk state (0.31 eV endothermic) most of defects can bind Li exothermically, and therefore stably w.r.t. clustering. The strongest binding site is at the zigzag edge, due to the presence of dangling bonds.[28] Our results suggest that Li can be stored in disordered graphene, which could possibly give rise to the capacity observed in some experiments.[17, 18] In order to achieve a high Li capacity for practical applications, one would need to fabricate highly defective graphene. This is in the contrary to the mainstream efforts to synthesize defect-free graphene,[29-32] but may be possible with amorphous graphene produced by irradiation.[33]

*Anchoring of other Li-adsorbing materials* (silicon,[34, 35] metal oxides,[36-38] etc.) to graphene surface should be mentioned, although we do not perform here any actual computations of specific systems. Not only the high surface-mass ratio but also the high conductivity of graphene could be utilized in this approach.[39] However, the clustering of Li-adsorbing materials could be a potential problem, similar to the reduction of hydrogen uptake induced by the clustering of hydrogen-adsorbing metals.[40-42]

*Chemical doping.*[43, 44] Since Li donates its 2$s$ electron to the matrix, an electron-deficient matrix, such as B-substituted C, could better accommodate for extra electrons. Figure 2 shows that, indeed, binding is stronger at B substitution site than on pristine graphene, while it is weaker at the electron-abundant N-substitution site. Besides, such dopants are inherent part of the matrix lattice, which eliminates the problem of dopant clustering. Therefore, highly B-doped graphene, or in other words, 2D C-B compound, should be a good candidate for Li storage. In fact, recent studies have confirmed that the Li storage can be enhanced by B doping.[44-46] Graphene can also be doped with other elements such as Si, P, and S. For comparison, the Li binding energies ($\varepsilon_{Li} - \varepsilon_{Li-M}$, where M = B, N, Si, P, S) are calculated, which are -0.88, 0.82, -0.40, -0.38, 0.21eV/atom, respectively. Clearly the B-doped graphene has the strongest binding with Li, suggesting a possibly highest capacity. In addition, only B and N dopants can keep the originally planar structure of graphene, while the other dopants are buckled by ~1.6 Å. The significant distortions imply the possible instability of these dopants. Moreover, solid experimental evidence of stable 2D C-Si, -P, and -S compounds are still lacking. We therefore focus on the C-B system. The experimentally available 2D compound with the highest B:C ratio is $^{2D}C_3B$, which has a 2D structure with C-hexagons connected by B atoms,[45-50] shown in Figure 3. The C$_3$B layers can be stacked up to form graphite-like 3D structure $^{3D}C_3B$, with weak van der Waals interactions between layers.[51, 52] In the following, we discuss the Li storage in the C$_3$B in some detail since it appears potentially interesting for anode applications.

The lithiation curves and atomic structures of the Li-saturated C$_3$B are shown in Figure 3. The corresponding atomic structures are shown in Figure S2. During lithiation, the $^{3D}C_3B$ preserves its layered structure but changes the stacking from AB order[53] to AA (every next layer is directly on top of the previous one). This behavior is similar to that of graphite, suggesting a small volume variation in discharge/charge cycles. The Li-saturated $^{3D}C_3B$ has all the hexagons occupied by Li except those composed entirely of C, resulting in the Li$_{1.5}$C$_3$B composition with a capacity of 857 mAh/g, which is 2.3 times greater than that of graphite. Though $^{2D}C_3B$ has both its sides exposed for adsorbing Li, fewer hexagons are occupied in the fully-lithiated state, which has the Li$_{1.25}$C$_3$B composition with a capacity of 714 mAh/g. Once again, we see that the 2D material does not necessarily have higher capacity than its corresponding 3D form, in spite of higher surface-to-mass ratio. The reason of the Li capacity reduction in the $^{2D}C_3B$ is similar to that of graphene: binding for surface adsorption is weaker than that for intercalation. For example, at Li:C$_3$B = 0.5 ($x$ = 0.75) this difference is $\varepsilon_{Li-^{2D}C_3B} - \varepsilon_{Li-^{3D}C_3B} = -1.20$ eV. On the other hand, the weaker binding to $^{2D}C_3B$ could turn beneficial for battery voltage: if used as the anode, the $^{2D}C_3B$ should yield higher average voltage than $^{3D}C_3B$ by 0.52 V. Taking the cathode half-cell voltage of 3.7 V (corresponding to the commercially used cathode material LiCoO$_2$),[3] the estimated energy densities for $^{2D}C_3B$ and $^{3D}C_3B$ are very close, 2121 and 2100 Wh/kg, respectively, both far surpassing that of graphite (1347 Wh/kg). It is further interesting to note that if only one side of $C_3B$ is allowed to adsorb Li, it could reach the same high capacity as $^{3D}C_3B$ ($Li_{1.5}C_3B$, 857 mAh/g), while also maintaining a voltage even higher than for both-sides lithiation (surpassing the $^{2D}C_3B$ by 0.25 V, with an energy density of 2760 Wh/kg), as shown in the SI. It suggests a superior anode could be made of $C_3B$ capped single-wall nanotubes or foams,[54] where the Li ions cannot penetrate through the tubes into the inner region[55] and thus are mainly adsorbed onto the exterior of tubes.

As discussed above, the enhanced Li storage in $C_3B$ results from the greater binding, $\varepsilon_{Li-C_3B}$. This strong binding is explained by the charge density difference between Li-saturated and pure $C_3B$, as shown in Figure 4. There are no valence electrons surrounding Li, indicating that Li is fully ionized. The electrons transferred from Li to $C_3B$ are mainly concentrated on B, filling the originally empty $p_z$ states of B. Due to the better accommodation of the transferred electrons $C_3B$ has a higher binding energy with Li than that of graphite.

In spite of significantly different binding energies $\varepsilon_{Li-M}$, the diffusion activation barriers for the Li ions in both matrices are similar. One of the diffusion mechanisms at high Li concentration is vacancy hopping, shown in Figure 4, which has a barrier of 0.40 eV, comparable with that in graphite 0.34 eV (using consistent calculations settings, shown in the SI). In reality, the diffusion is more complicated since the large size anode inevitably contains defects which impact the diffusivity in different ways. For example, the Li transport perpendicular to the basal plane of graphite is facilitated by the defects, whereas the diffusion parallel to the plane is limited by the defects.[56] The influence of the defects on Li diffusivity deserves further study. Although the pristine $C_3B$ sheet is a semiconductor with a band gap of ~0.5 eV,[57] it becomes metallic during lithiation, as demonstrated by the electronic density of states plot in the Figure 4. The similar ionic and electronic conductivity between $C_3B$ and graphite should give comparable discharge/charge rates for the battery. Overall, $C_3B$ has a larger capacity and similar power density compared to graphite, but somewhat lower voltage as a consequence of larger $\varepsilon_{Li-M}$.

In summary, although nanomaterials provide more free surfaces for adsorption compared with bulk materials, they might suffer from the weakened adsorbate-adsorbent binding, which could lead to the adsorbates clustering and a decreased adsorbate capacity. This conclusion is exemplified by Li storage in graphene, where Li phase separation results in significant capacity limitations (down to zero for pristine monolayer graphene). The feasibility of modifying graphene to store Li more efficiently is discussed, including its doping, and leading one to stoichiometric 2D compound $C_3B$ as a promising electrode material. Its capacity is about twice larger than graphite, with comparable power density and small volume variation during discharge/charge cycles. Our results help to clarify the fundamentals of Li storage in low-dimensional materials, and shed light on the rational design of nano-architectures for energy storage.

METHODS

The structures are relaxed and the total energies of the systems are calculated by density functional theory (DFT) with generalized gradient approximation (GGA). Although the DFT-GGA methods have been widely used to study the Li-ion battery electrodes and achieved good agreements with experiments,[3, 24] one has to be aware that the approximate functional suffers from the "delocalization error" and overestimated the polarizability and the binding energy of the charge transfer complex.[58] Hybrid functional might help to obtain more accurate energetics,[58] while it is too costly for the large systems addressed in this work and unlikely to significantly alter the main conclusions which are based on the ground state properties. To determine the ground state properties, each Li site in the matrix is assigned with one occupation variable $\sigma_i$, which is +1 if occupied by the Li or -1 if empty. Within the CE formalism,[22] the total energy of the system can be expanded over the 'clusters' of sites: $E_{tot} = C_0 + \sum_{ij}C_{ij}\sigma_i\sigma_j + \sum_{ijk}C_{ijk}\sigma_i\sigma_j\sigma_k + \ldots$ The coefficients $C$ are determined with ATAT code,[23] by fitting the energies to the direct DFT-computed values of different configurations. After getting a representative set of clusters and the corresponding coefficients, the energy of any given lattice configuration can be directly obtained using the above equation without DFT computation. The ground state structure and energy can thus be identified from the complete set of all possible configurations for the chosen supercell size. The details of the computations can be found in the SI.

ASSOCIATED CONTENT

**Supporting Information**. Details on computational methods, the atomics structures of lithiated $C_3B$, and Li diffusion in graphite. This material is available free of charge via the Internet at http://pubs.acs.org.

AUTHOR INFORMATION

**Corresponding Author**

*Email: biy@rice.edu

ACKNOWLEDGMENT

This work is supported by the Honda Research Institute USA. The computations were performed at (1) the NICS Kraken, funded by the NSF grant OCI-1053575, (2) the NERSC Hopper, supported by the DOE grant DE-AC02-05CH11231, and (3) the DAVINCI, funded by the NSF grant OCI-0959097. The authors thank Dr. Hoonkyung Lee and Dr. Xiaolong Zou for valuable discussions.

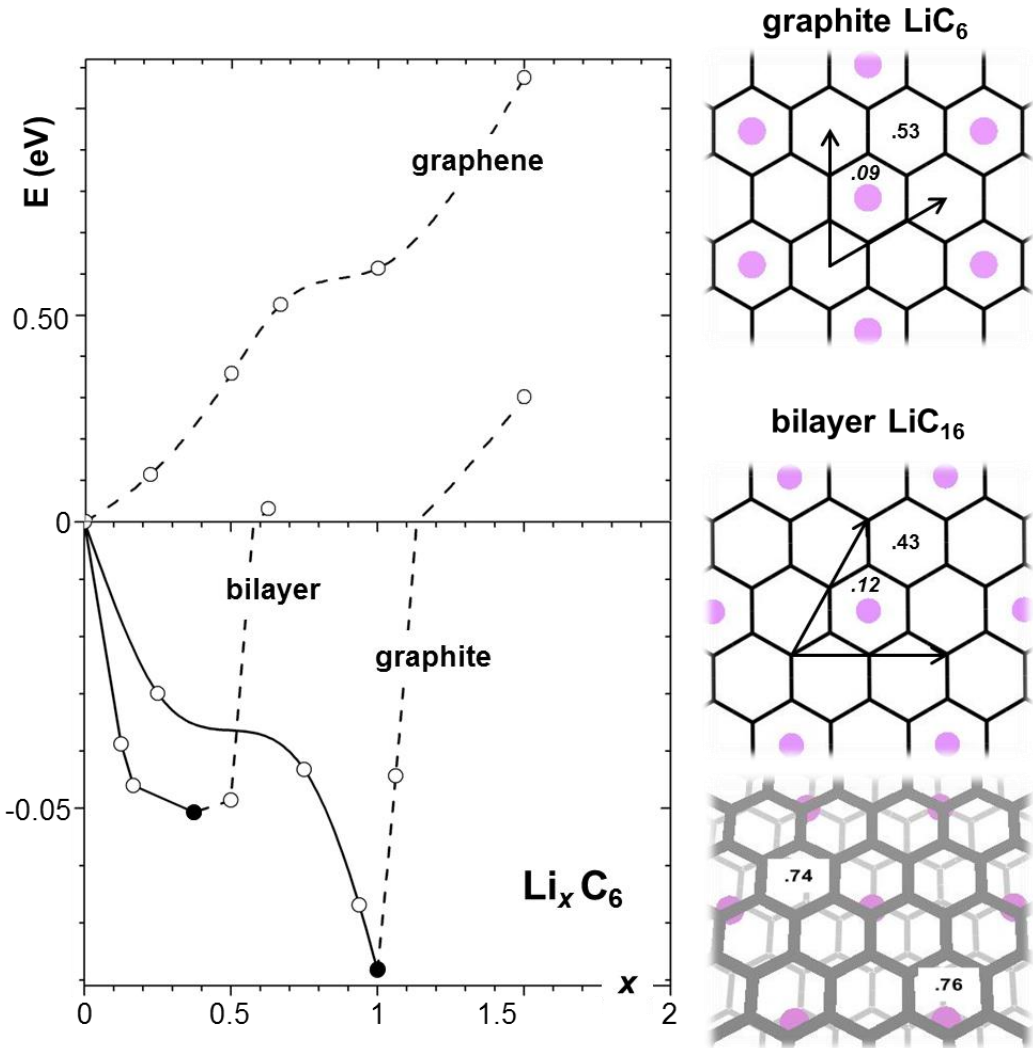

**Figure 1**. Left: lithiation energy (defined in Equation 1) as a function of Li amount *x* in different C matrices: graphite, graphene, and bilayer graphene. The positive and negative binding energy domains are shown in different scales. Right: atomic structures of Li-saturated graphite and bilayer graphene (top and perspective views). The Li is represented by balls and the C matrix by sticks. The arrows indicate the primitive cell vectors. The numbers are the energies of adding (*removing,* in italics) a Li atom to (*from*) the empty (*Li-filled*) hexagons, all in eV. For the bilayer, the energies of Li intercalation/deintercalation between the layers are shown in the top view, and those for Li adsorption/desorption at the exterior are marked in the perspective view.

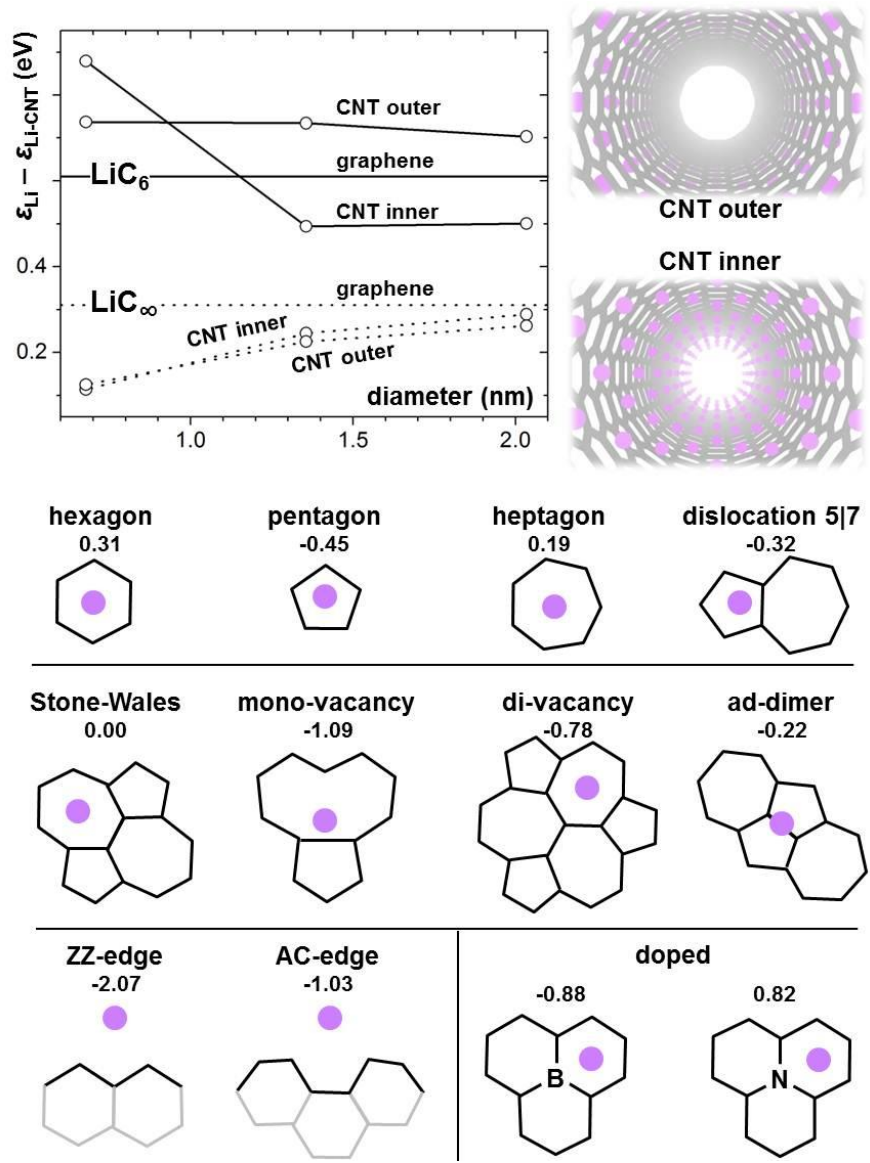

**Figure 2**. Energies and structures of Li adsorbed on carbon nanotubes (CNT) and defects. The balls show Li, and the sticks represent C lattice. The numbers are calculated as $\varepsilon_{Li} - \varepsilon_{Li-M}$, in eV. The energy of a single Li atom adsorption on pristine graphene (hexagon) is also shown for comparison. The plots show $\varepsilon_{Li} - \varepsilon_{Li-CNT}$ as a function of diameter for (5, 5), (10, 10), and (15, 15) CNTs, for adsorption on inner or outer surfaces, at high (LiC$_6$, solid lines) or low ((LiC$_\infty$, dashed lines) concentrations.

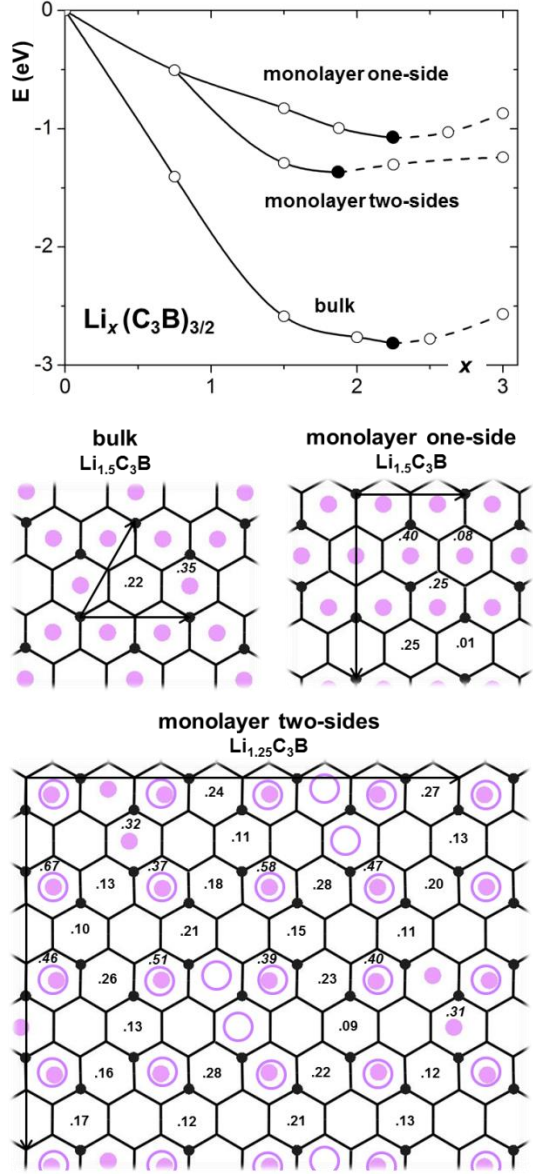

**Figure 3**. Lithiation energy (defined in Equation 1) as a function of Li amount in $C_3B$ matrix, and atomic structures of Li-saturated bulk and monolayer $C_3B$. The matrix is shown by sticks with B-substitutions marked by black balls, and the Li ions are represented by the pink circles (solid for top/front and unfilled for bottom/back sides of monolayer $C_3B$). The arrows indicate the primitive cell vectors. The numbers are the energy change upon adding (or removing) a Li ion to (from) the empty (Li-filled) hexagons. For monolayer $C_3B$, the energies are shown for Li adsorption/desorption on the top surface.

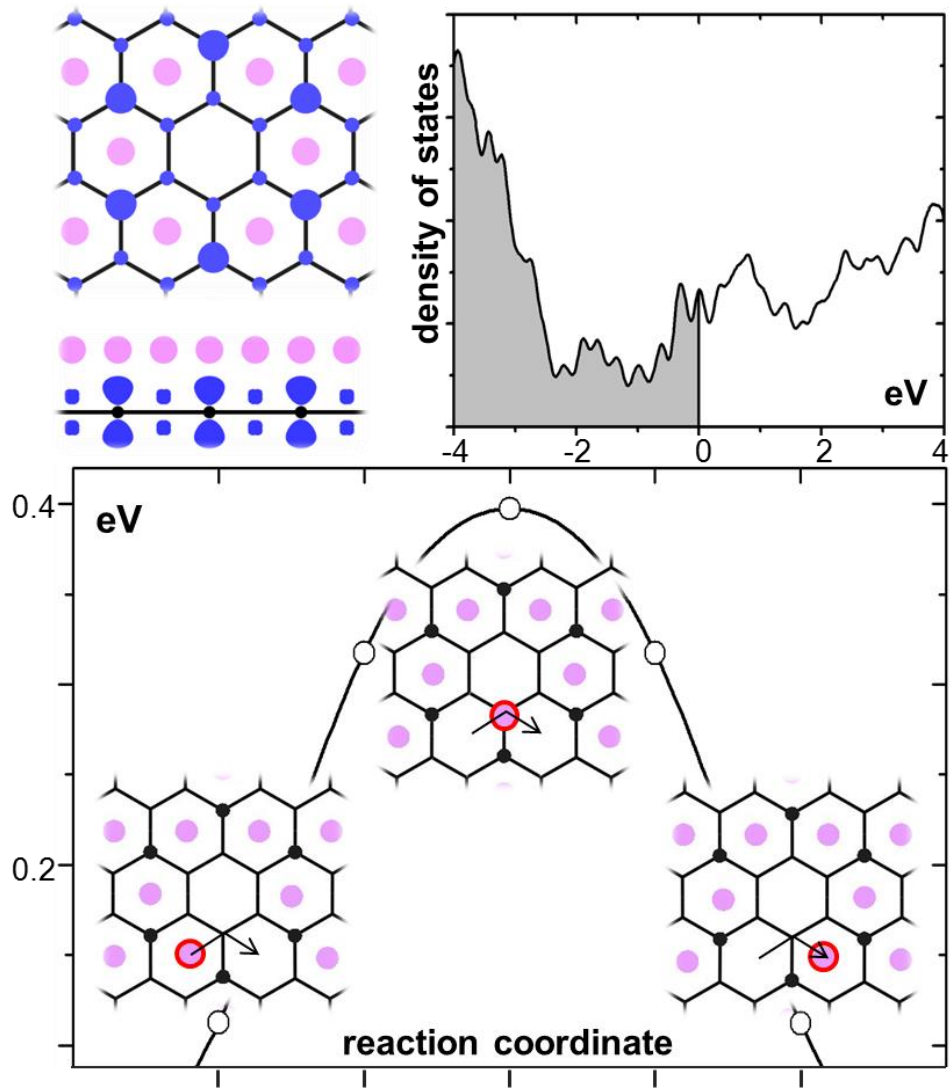

**Figure 4.** Top left: charge density difference between Li-saturated and Li-free $^{3D}C_3B$; the electron accumulation region is shown by blue isosurfaces, for both top and side views. The matrix is shown by sticks with B-substitutions marked by black balls, and the Li ions are represented by the pink balls. Top right: electronic density of states of Li-saturated $^{3D}C_3B$, with Fermi level at 0 eV. Bottom: energies and pathways of vacancy-hopping diffusion in Li-saturated $^{3D}C_3B$. The insets from left to right show the initial, transition state, and final structures, respectively.